\pdfoutput=1 
\documentclass[%
 aip,%
rsi,
 amsmath,amssymb,%
 reprint,%
 groupedaddress,
]{revtex4-1}

\usepackage{graphicx}
\usepackage{dcolumn}
\usepackage{bm}
\usepackage{enumitem}
\begin{document}

\preprint{AIP/123-QED}

\title 
{A new energy spectrum reconstruction method for Time-Of-Flight diagnostics of high-energy laser-driven protons}

\author{G. Milluzzo$^{1,2,3}$}
\author{V. Scuderi$^{4,2}$}
\author{A. Alejo$^{1}$}
\author{A. G. Amico$^{2}$}
\author{N. Booth$^{5}$}
\author{M. Borghesi$^{1}$}
\author{G.A.P. Cirrone$^{2}$}
\author{G. Cuttone$^{2}$}
\author{D. Doria$^{6,1}$}
\author{J. Green$^{5}$}
\author{S. Kar$^{1}$}
\author{G. Korn$^{4}$}
\author{G. Larosa$^{2}$}
\author{R. Leanza$^{2}$}
\author{D. Margarone$^{4}$}
\author{P. Martin$^{1}$}
\author{P. McKenna$^{7}$}
\author{G. Petringa$^{2,3}$}
\author{J. Pipek$^{2}$}
\author{L. Romagnani$^{8,1}$}
\author{F. Romano$^{9,2}$}
\author{A. Russo$^{2}$}
\author{F. Schillaci$^{4,3}$}

\affiliation{$^{1}$School of Mathematics and Physics, Queens University Belfast, Belfast, U.K.
}%
\affiliation{$^{2}$INFN-Laboratori Nazionali del Sud (LNS-INFN), Via S Sofia 62, Catania, Italy
}%
\affiliation {$^{3}$ Physics and Astronomy Department, University of Catania, Catania, Italy}

\affiliation{$^{4}$Institute of Physics ASCR, v.v.i. (FZU), ELI-Beamlines Project, Prague, Czech Republic
}%

\affiliation{$^{5}$Central Laser Facility, STFC Rutherford Appleton Laboratory, Didcot OX11 0QX, U.K
}%

\affiliation{$^{6}$Extreme Light Infrastructure - Nuclear Physics (ELI-NP), Horia Hulubei
Institute for Nuclear Physics (IFIN-HH), Reactorului Str., 30, Magurele
Campus, Bucharest}

\affiliation{$^{7}$Department of Physics, SUPA, University of Strathclyde, Glasgow G4 0NG, U.K
}%

\affiliation{$^{8}$LULI, Ecole Polytechnique, CNRS, CEA, UPMC, 91128 Palaiseau, France
}%

\affiliation{$^{9}$National Physical Laboratory, Hampton Rd, Teddington, Middlesex, UK
}%




\date{\today}

\begin{abstract}
The Time-of-Flight (ToF) technique coupled with semiconductor-like detectors, as silicon carbide and diamond, 
is one of the most promising diagnostic methods for high-energy, high repetition rate, laser-accelerated ions allowing a full on-line beam spectral characterization.

\noindent A new analysis method for reconstructing the energy spectrum of high-energy laser-driven ion beams from TOF signals is hereby presented and discussed. The proposed method takes into account the detector's working principle, through the accurate calculation of the energy loss in the detector active layer, using Monte Carlo simulations.
\noindent The analysis method was validated against well-established diagnostics, such as the Thomson Parabola Spectrometer, during an experimental campaign carried out at the Rutherford Appleton Laboratory (RAL, UK) with the high-energy laser-driven protons accelerated by the VULCAN Petawatt laser.

\end{abstract}

\maketitle


\section{\label{sec:level1} Introduction } 

Over the last decades, the interest towards innovative particle acceleration techniques, alternative to conventional methods, has led to a growing effort in the study of high power laser-plasma interactions \cite{Macchi2010, Macchi2013} 
as sources of a varied range of particles. 
In particular ultra-short multi-MeV laser-accelerated ions, if well controlled, could provide a promising alternative tool for dose delivery during radiobiological irradiations as well as, in a future perspective, for clinical treatments (hadrontherapy) \cite{Bulanov2014,Busold2014, Cirrone2015, Masood2017, Margarone2018, Higginson2018}.
In this framework, the development of novel instrumentation aimed at measuring with good accuracy ion beam characteristics, such as energy distribution, flux and shot-to-shot reproducibility, represents a crucial step towards obtaining controlled beams, exploitable for applications.
Well established detectors typically used in laser-driven acceleration experiments are for instance radiochromic films (RCF), nuclear track detectors (CR-39), image plates 
\cite{Bolton2014}.
 They have been widely used for beam diagnosis with low-repetition rate laser systems operating in single-shot mode and all require some form of processing after the exposure.
Nevertheless, thanks to ongoing advances in laser technologies, the acceleration of high-energy ions at high repetition rate is becoming possible and therefore 
the real-time diagnosis  of laser-generated particles is a key point for the shot-to-shot monitoring of the beam parameters beam needed for applications.
The Time-of-Flight (ToF) technique, so far used as on-line diagnostics for low-energy laser accelerated proton and ion beams, enables measurement of the ion maximum energy (cut-off), spectrum and flux \cite{Margarone2008, Margarone2011}.
As discussed in literature \cite{Margarone2011, Margarone2012, Bertuccio2013, Marinelli2013, Musumeci, Milluzzo_1, Scuderi}, detectors as Faraday Cups (FC), ion collectors (IC) and semiconductor-like detectors such as diamond and silicon carbide, are typically placed at a finite distance from the target to measure TOF of accelerated particles.
The TOF signal generated in such devices results from the contribution of the different ion species incoming with a specific TOF and, as a consequence, kinetic energy. 
According to the approach discussed in \cite{Miotello, Kelly,Krasa2013, Milluzzo_2}, TOF signals can be typically described as a convolution based on the so-called  \emph{Maxwell-Boltzmann Shifted} (MBS) functions defined for each ion species.
The signal deconvolution by means of the MBS functions also allows investigating plasma parameters, such as the ion temperature and shift velocity, defined as the additional component to be considered due to the plasma center-of-mass velocity, as it is reported in \cite{Kelly}.
When a single ion species, e.g. protons, is detected, the signal amplitude can be directly converted in an energy spectrum, considering the detector response to the incoming radiation. 
IC-FC and semiconductor detectors exhibit extremely different responses to charged particle radiation thus different approaches must be followed for the energy spectrum reconstruction from TOF signals.
For instance, ion signal amplitudes generated in IC or FC devices depend on the collected charge, thus  the absolute number of incoming particles can be  obtained. 
On the other hands, for semiconductor-like detectors whose response depends on the energy deposited within the active detector layer and on the detector electron-hole pair energy creation, the signal amplitude needs to be converted. 
Nevertheless the higher signal-to-noise ratio of semiconductor-like detectors compared to FC and IC detectors, coupled with their good performances, in terms of rise time and time resolution \cite{DeNapoli2009, Randazzo2012} make silicon carbide and diamond detectors particularly suitable for the TOF diagnosis of high-energy particles \cite{Scuderi}.\\
A new analysis processing method is here proposed for reconstructing the ion energy spectrum from TOF signals generated by high-energy, laser-accelerated ions and acquired with semiconductor-like detectors. 
Such analysis procedure was successfully applied for the energy spectrum reconstruction of high-energy protons accelerated from the Vulcan PW laser system in a wide energy interval up to 30 MeV.

\section{\label{procedure} A new procedure for proton energy spectrum reconstruction}

As it is reported in \cite{Milluzzo_1,Milluzzo_2,Scuderi} a typical TOF signal displays the temporal evolution of the radiation and particles reaching the detector after being emitted from the laser-irradiated target:
UV/X-rays promptly emitted when the laser hits the target surface generate the \emph{trigger} signal in the detector, i.e. the so-called photo-peak which is typically followed by a tail originated from fast electrons. Protons and heavier ion species, according to their energy and flight path, follow at longer TOF with a typical narrow peak (fast protons) and a broad signal (slow protons and ions).  
The kinetic energy of a given ion species, for instance protons, is then calculated from the measured proton arrival time with respect to the photo-peak ($t_{ion}$), knowing the flight path L and considering the time needed from UV and X-rays to travel the distance from the source to the detector ($t_{ph}$): 

\begin{equation}
TOF=t_{ion}+t_{ph}  \qquad t_{ph}=\frac{L}{c}
\end{equation}
where c is the speed of light and TOF is the time of flight corrected for the photon flight path. The proton kinetic energy can be obtained by the well-known relativistic definition:
\begin{equation}
\label{Ekin}
E_{kin}=(\gamma-1) M_p c^2 \qquad \gamma=\frac{1}{\sqrt{(1-\beta^2)}}\qquad \beta=\frac{L}{cTOF}
\end{equation}
where $M_p$ is the rest mass of proton or ion. \\
The absolute number of particles as well as the energy spectrum can be extracted from the signal amplitude, knowing the detector characteristics.
As it is well known, the charge collected in semiconductor-like detectors due to \emph{N} incident particles, depositing energy \emph{E} in the active layer, can be expressed by: 
\begin{equation}
Q=\frac{eNE}{E_g}
\label{charge}
\end{equation}
where $E_g$ is the electron-hole pair energy creation, i.e. 7.78 eV and 13 eV respectively for silicon carbide (SiC) and diamond.

The detector can be directly connected to a fast oscilloscope for the acquisition, with a typical time sampling of the order of 10$^{-10}$ s. 
From eq.\ref{charge}, the following expression relating the ion current \textbf{i}(t) and the energy spectrum \emph{dN/dE} for a given ion species, can be obtained:

\begin{equation}
\frac{dN}{dE}=\frac{\epsilon_g i(t)}{eE^2}(-\frac{1}{2}t-\delta t)
\label{spectrum}
\end{equation}

\noindent where $i(t)=V(t)/R$ with V(t) being generally the measured signal amplitude or the amplitude of the MBS function describing the original TOF signal and obtained from the signal deconvolution, \emph{R} the total resistivity of the detector readout (including termination on the oscilloscope), \emph{t} corresponds to the measured TOF corrected with the light traveled distance \cite{Milluzzo_1}, $\delta$t is the oscilloscope time sampling (negligible by using fast scopes) and \emph{L} is the flight path.
Integrating the equation \ref{spectrum} the number of particles of a given species impinging on the detector can be extracted. \\
Considering the kinetic energy of the incident particles, two different regimes can occur:

\begin{itemize}
\item \emph{Case 1}. Particles having an incident kinetic energy such that they stop within the detector active thickness. In such case, the energy spectrum is reconstructed from eq. \ref{spectrum} with the parameter \emph{E} corresponding to the kinetic energy obtained with eq. \ref{Ekin}. 

\item \emph{Case 2}. Particles having a sufficient kinetic energy to traverse the detector thickness, releasing only a fraction of their incident energy inside the active layer. In such case, the energy loss in the detector, corresponding to the incident energy measured with TOF technique, needs to be calculated. 

\end{itemize}
So far proton energy spectra have been measured from TOF signals for low-energy protons when \emph{Case 1} occurred as it is reported in literature \cite{Margarone2011, Bertuccio2013,Marinelli2013}. 
For \emph{Case 2}, involving proton energies from few up to hundred MeV, a reliable reconstruction method is required to make the TOF technique an established diagnostic tool in view of its use in experiments with high-energy laser-driven ions. \\
\noindent The approach proposed here takes into account the mechanism of signal formation in the specific detector employed in order to reconstruct the absolute number and energy spectrum of the impinging particles. 
Monte Carlo simulations are used to calculate the energy loss inside the detector active layer corresponding to the incident kinetic energy measured with the TOF technique. 

The detector is simulated, using the Monte Carlo Geant4 \cite{Allison} code, in terms of material, density and thickness and a point-like parallel beam is used as input source assuming a uniform energy distribution within the energy interval of interest. The particles generated in the Geant4-based application are tracked and information, such as kinetic energy and position, is retrieved along the track.
The energy deposited from each simulated particle within the detector thickness is then retrieved and can be related to the corresponding incident kinetic energy.
In such way a correlation between the kinetic energy and the energy loss can be obtained and used to extract from the experimental data the energy spectrum in the selected energy interval. 

Such method was used to analyze the TOF signals acquired in different laser-acceleration experiments using laser systems from few TW up to PW power accelerating protons with energies ranging from few MeV up to 30 MeV \cite{Giuffrida, Milluzzo_1, Margarone2016}. The procedure was validated against other well-established diagnostics and optimized adapting it according to the different experimental conditions and purposes.

\begin{figure}
\includegraphics[width=200pt,height=150pt]{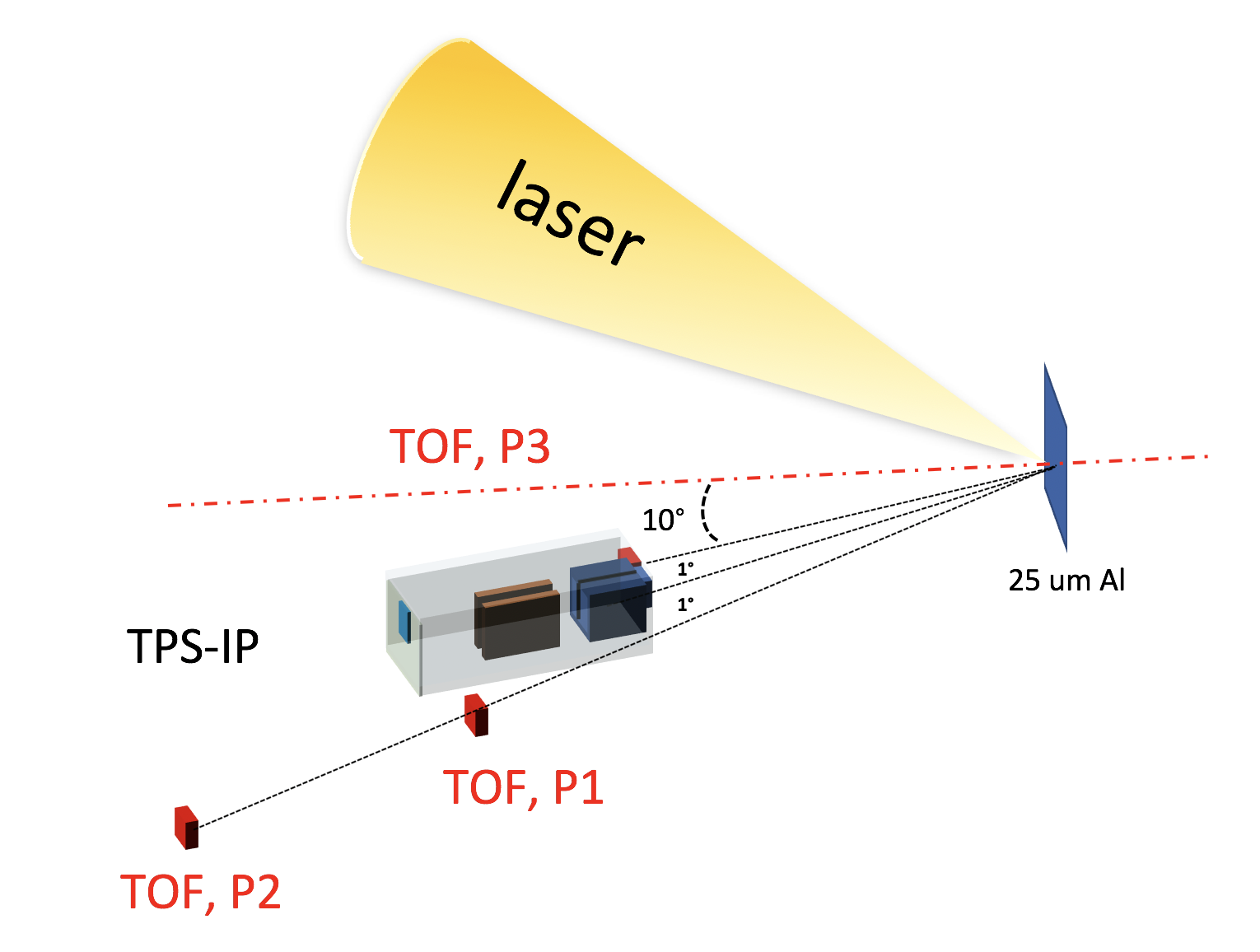}
\caption{Experimental setup: the laser incidence angle (laser- target normal) was about 20$^\circ$ and the TPS-IP was placed at about 10$^\circ$. The three positions of the diamond detector used for TOF measurements (TOF) are also indicated: P1 and P2 have respectively a 2.35 m and 4.10 m flight path. P3 indicates the position of the diamond detector placed alongside the TPS pinhole at about 1.2 m from target. }


\label{setup}
\end{figure}

In particular, the TOF technique employing diamond detectors was used to investigate proton acceleration during a recent experimental campaign carried out using the VULCAN PW laser at the Rutherford Appleton Laboratory. 
A laser pulse of wavelength 1.054 $\mu$m, duration of $\sim$ 700 fs and energy up to $\sim$ 400 J on target, was focused onto a 25 $\mu$m-thick Al target leading to the acceleration (from surface contaminants) of protons and light ions, such as carbon and oxygen. 
A Thomson Parabola Spectrometer (TPS) coupled with Image Plates was placed in the backward direction at about 1.2 m  from target, separating the ion species according to charge-to-mass ratio and providing the ion energy cut-off and spectra measurements using the analysis method and the calibration reported in \cite{Alejo,Doria2015}. 
A 100 $\mu$m thick polychrystalline diamond detector was placed at the target front side at about 2.35 m (P1) from the target location, with an applied voltage of 200 V. The experimental setup is shown in FIG. \ref{setup}. 

\begin{figure}[h!]
\centering
\includegraphics[width=220pt,height=150pt]{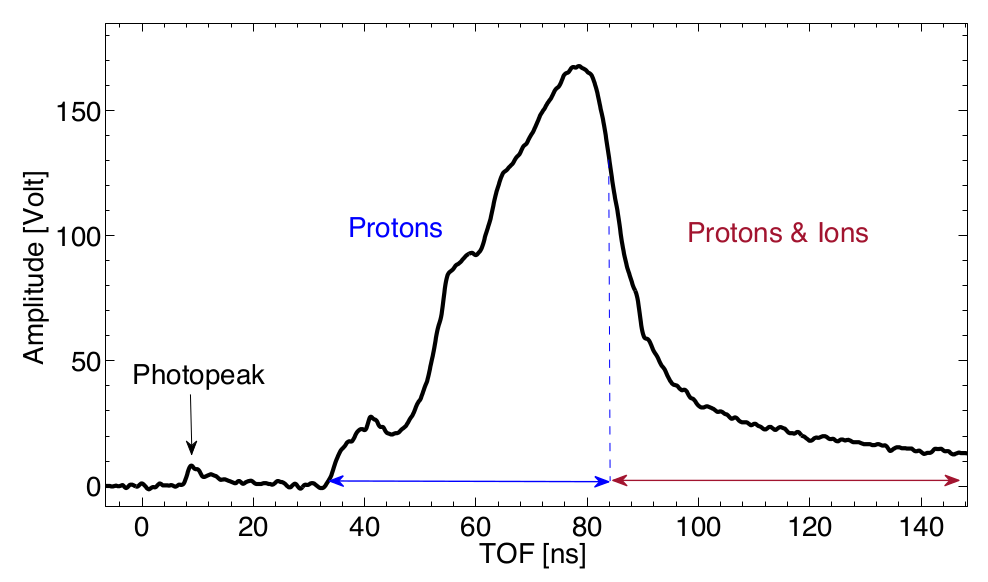}
\caption{TOF signal acquired with a 100 $\mu$m thick polychristallyne diamond detector during the experiment carried out with Vulcan Petawatt.}
\label{intervallo}
\end{figure}

As one can see in FIG. \ref{intervallo} the acquired TOF signal is composed of a small photopeak (7-10 ns) generated by plasma soft X-ray/ultraviolet (XUV) emission and a broad peak (32-140 ns) resulted as the sum of protons and other contaminants.
According to TPS measurements, carbon and oxygen ions in different charge states are accelerated together with protons, with a maximum energy per nucleon not exceeding 4 AMeV, which for carbon ions corresponds to a TOF@2.35 of about 80 ns.
As a consequence, the TOF signal in the time interval 32-80 ns is uniquely originated from protons 
and the energy distribution can be reconstructed following the developed procedure.

\begin{figure}[h!]
\centering
\includegraphics[width=200pt,height=140pt]{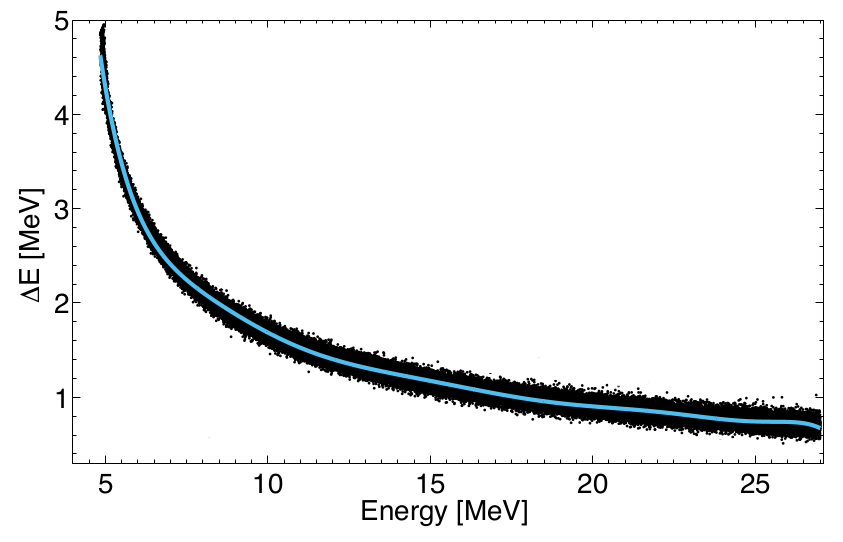}
\caption{Energy loss ($\Delta E$) simulated with Geant4 for protons with energies ranging from 5
and 27 MeV in 100 $\mu$m-thick diamond.}
\label{delta}
\end{figure}

According to the calculation performed with the LISE++ analytical code \cite{lise} protons with energy higher than 5 MeV traverse the detector thickness (case 2). 
A simulation reproducing the experimental condition, namely protons with energies higher than 5 MeV impinging on a 100 $\mu$m thick diamond detector, was performed to retrieve the corresponding energy loss and reconstruct the energy spectrum.
The energy loss ($\Delta$E) as a function of the incident kinetic energy is thus obtained by means of the simulations as it is shown in FIG. \ref{delta}. 



As it is expected, the energy loss calculation takes into account the energy straggling effect, originating from stochastic fluctuations in the energy loss of protons. 
The FWHM of such energy loss distribution, typically of the order of hundreds of KeV, was considered in the analysis as a contribute to the uncertainty on the energy loss calculation.

A function relating the kinetic energy and the average energy loss was obtained through a polynomial fit (light blue line in FIG. \ref{delta}). 
The parameters resulting from the fit were then used to associate an energy loss to the kinetic energies calculated from the measured TOF values. 

\begin{figure}[h!]
\centering
\includegraphics[width=200pt,height=140pt]{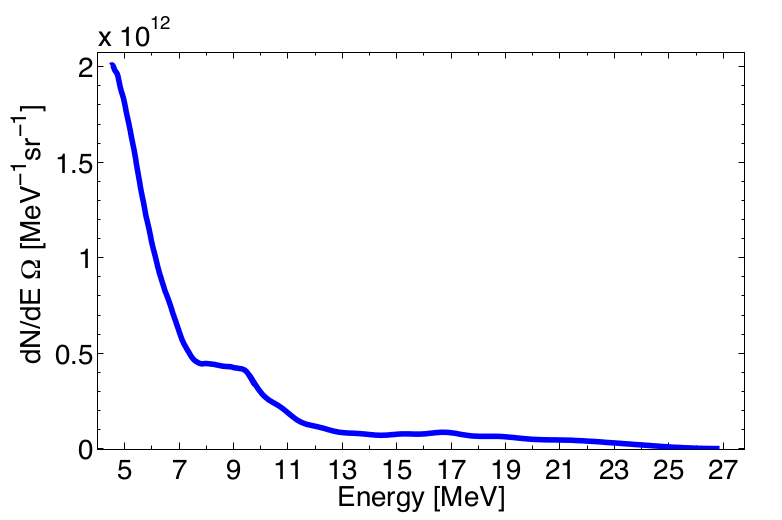}
\caption{Proton energy spectrum reconstructed following the proposed procedure from the TOF signal shown in FIG. \ref{intervallo}.}
\label{espectrum}
\end{figure}

The energy spectrum obtained in the range 5-27 MeV is shown in FIG. \ref{espectrum}. Integrating the energy distribution a number of protons per solid angle of (5 $\pm$ 1) $\times$ $10^{12}$ protons/sr was estimated. 
The uncertainty on the number of proton per steradian is calculated applying the error propagation on eq. \ref{spectrum} and considering the statistical uncertainty. An uncertainty on the energy loss calculation is of the order of 10$\%$ and was evaluated  taking into account the energy straggling effect. Uncertainty on the solid angle of the order of 0.1$\%$ was considered. \\
As it is shown in figure \ref{intervallo}, besides protons, other ion species can be accelerated from the target and  will overlap in the detector signal with protons having the same TOF. 


In FIG. \ref{intervallo} this is the case for TOF$>$ 80 ns, where the TOF signal results from the overlap of different ions (carbon,oxygen and protons) and the single species contribution as well as the corresponding energy distributions cannot be discriminated. \\
Metallic or plastic absorbers with different thicknesses are typically placed in front of the detector to select the high-energy proton component and filter the ions and the high-flux low-energy protons, which could saturate the detector.

\begin{figure}[h!]
\centering
\includegraphics[scale=0.4,keepaspectratio]{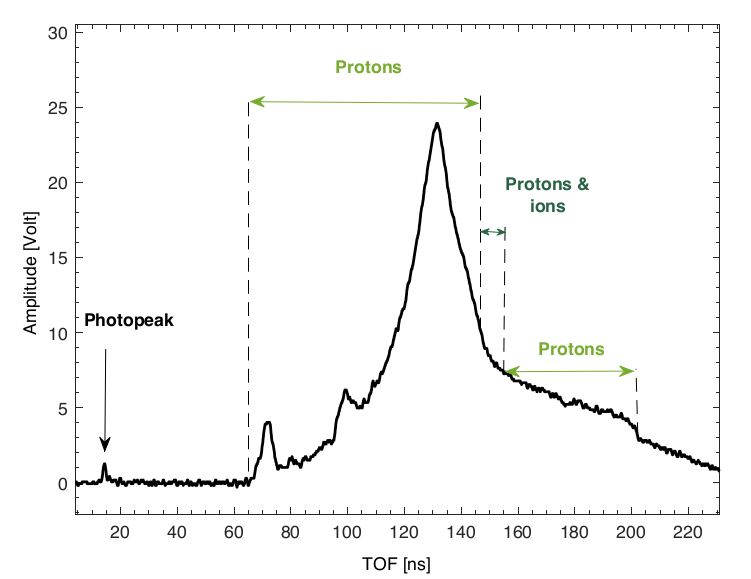}
\caption{TOF signal acquired in the backward direction with the 100 $\mu$m thick diamond detector for flight path = 4.10 m. A 50 $\mu$m Al foil was used to stop heavy ions. Proton and ion contributions in the three different time intervals are shown.}
\label{shot25}
\end{figure}

With an appropriate filter, a TOF signal uniquely due to protons can be obtained and the proton energy spectrum can be then extracted.  
FIG. \ref{shot25} shows a TOF signal acquired with a 100 $\mu$m thick diamond detector placed at about 4 m (P2) from target in the backward direction (FIG. \ref{setup}).
A 50 $\mu$m Al foil absorber was used in front of the detector.  According to the calculations performed with the code LISE++, such absorber stops protons, carbon and oxygen ions with energies up to 2.23 MeV, 3.7 AMeV and 4.2 AMeV respectively. 


The TOF values corresponding to such energies at 4.10 m for protons, carbon and oxygen ions are respectively 200 ns, 153 ns and 142 ns. 
Coupling together these considerations and the maximum ion energies measured with the TPS of about 4 MeV/n, which corresponds to a TOF of 148 ns@4.10 m, three time intervals can be identified in the TOF signal shown in FIG. \ref{shot25}: 1) 65-148 ns, the signal can be attributed uniquely to higher-energy proton contribution; 2) 148-153 ns, the signal results from the overlap of H$^+$, C$^{12}$  and O$^{16}$ ions in different charge states; 3) 154-200 ns, the signal originates only from low-energy proton component (from 2.23 MeV up to 3.7 MeV).

\begin{figure}
\includegraphics[width=200pt,height=140pt]{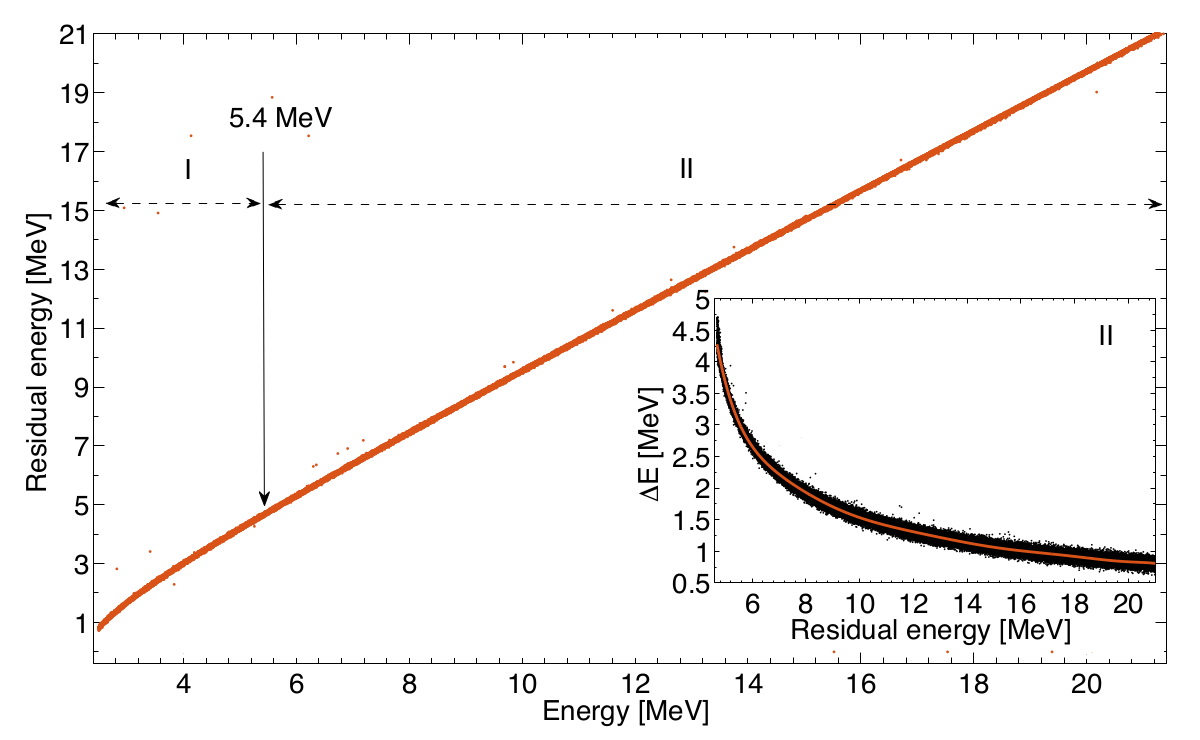}
\caption{Residual energy as a function of incident kinetic energy in the energy interval between 2.2 MeV and 21 MeV estimated with MC simulation.  }
\label{eres}
\end{figure}

\begin{figure}
\includegraphics[width=200pt,height=140pt]{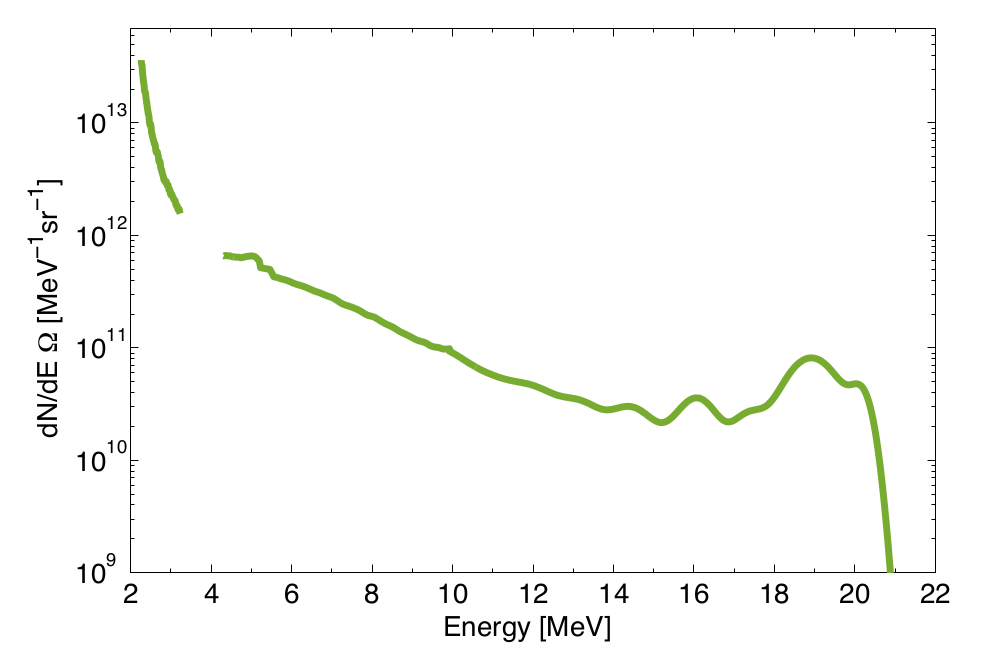}
\caption{TOF signal acquired in backward direction with the 100 $\mu$m thick diamond detector, flight path = 4.10 m. A 50 $\mu$m Al foil was used to stop heavy ions. Proton and ion contributions in 3 different time intervals are shown.}
\label{energy25}
\end{figure}

\begin{figure}[h!]
\includegraphics[width=180pt,height=140pt]{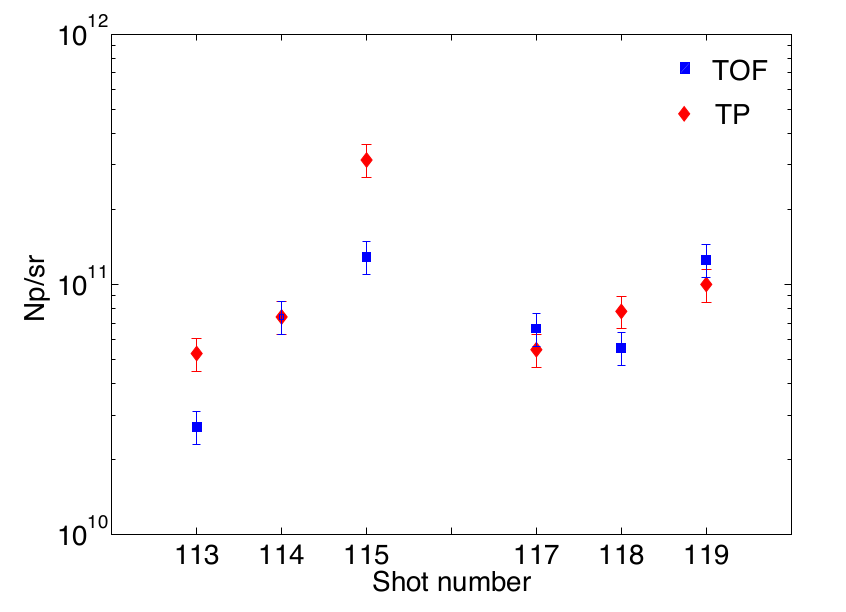}
\caption{Number of protons per solid angle in the energy range 13-18 MeV reconstructed with TOF technique (blue solid squares) and TPS images (red solid diamonds) for 6
consecutive shots.}
\label{Confronto}
\end{figure}

\begin{figure}[h!]
\includegraphics[width=180pt,height=140pt]{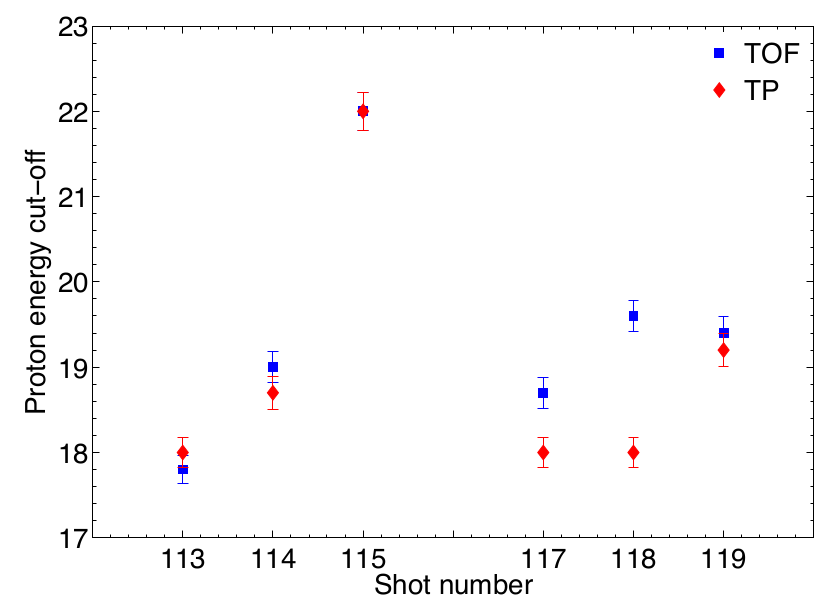}
\caption{Proton energy cut-offs measured with TOF technique (blue solid squares) and TPS images (red solid diamonds) for 6 consecutive shots.}
\label{ConfrontoB}
\end{figure}

In region 2)  H$^+$, C$^{12}$ and O$^{16}$ contributions cannot be disentangled from the TOF signal, since the signal is a convolution of different ion species and charge states. 
The proton energy distribution was therefore extracted in the regions 1) and 3) following the approach described earlier. 
In the presence of absorbers, Monte Carlo simulations are also needed to estimate, for particles traversing the absorber, the fraction of the incident energy lost within the absorber thickness and the residual energy in the detector thickness. 
Two cases can occur:

\begin{enumerate}[label=\Roman*.]
\item  Particles with residual energy after the absorber insufficient to traverse the detector thickness. These particles stop within the detector active layer and the variable \emph{E} in equation \ref{spectrum} corresponds to the particle residual energy.
\item  Particles with residual energy after the absorber sufficient to traverse the detector thickness. Considering the residual energy after the absorber, the corresponding energy released within the detector thickness has to be calculated with the help of Monte Carlo simulations.
\end{enumerate}
FIG. \ref{eres} shows the residual energy of protons traversing a 50 $\mu$m Al foil simulated with Geant4 as a function of the incident proton energy in the energy range from 2.2 MeV up to 21 MeV for the shot shown in FIG. \ref{shot25}. 
The two regions I and II are indicated in FIG. \ref{eres}. The proton energy corresponding to the transition between the two cases, i.e. 5.4 MeV, is also shown. 
Two different simulations have been performed for the two regions, according to the value calculated with the LISE++ code: 
for protons with an incident energy ranging from 2.2 MeV up to 5.4 MeV (region I), the residual energy after the filter is calculated by means of simulations, and can be directly used for the energy spectrum reconstruction; on the other hand for protons in region II, the energy loss within the detector thickness as a function of the residual energy has to be calculated (inset in FIG. \ref{eres}) similarly to the case shown in FIG. \ref{delta}.
The energy distribution corresponding to the TOF signal in FIG. \ref{shot25} is shown in FIG. \ref{energy25}: the gap between 3.7 MeV and 4 MeV corresponds to the region 2) in FIG. \ref{shot25}, where carbon ions contribute to the TOF signal besides protons.
\\ The new analysis method presented in this work was also used to analyze TOF data acquired with the diamond detector placed alongside the TPS (backward direction) at the same distance (1.22 m) and at about the same angle ($\sim$ 1$^\circ$) for several shots as it is shown in FIG. \ref{setup}. This allowed validating the TOF procedure through a direct comparison between the two diagnostics. 
FIG. \ref{Confronto} shows the number of protons per solid angle (Np/sr) obtained with the TOF developed procedure and with the TPS in the energy interval ranging between 13 MeV
and 18 MeV, for 6 consequtive shots. The energy range corresponds to the energy region common to all the shots analyzed.
A similar comparison between the maximum proton energies measured with the two diagnostics has been also performed for the the same shots, as it is shown in FIG. \ref{ConfrontoB}.
The TPS uncertainty was estimated to be about 15$\%$, considering the statistical and the calibration uncertainities. The results in FIG. \ref{Confronto} and FIG. \ref{ConfrontoB} confirm the good agreement between the two diagnostics, both sensitive to shot to shot fluctuations, indicating the good reliability of the developed TOF analysis method and its potential in the reconstruction of high-energy ion spectra with a good accuracy.
\\
\\

\section{Conclusions}

An analysis procedure, which allows converting the TOF signals measured with diamond detectors in energy distribution for a given ion species 
is here proposed. It takes into account of the mechanism for signal formation in semiconductor-like detectors, required to extract the energy spectrum.
The method hereby described was validated 
during an experimental campaign carried out at the Vulcan PW laser facility (RAL) and the obtained energy spectra are here shown together with a direct comparison with the measurement performed with a TPS, taken as a reference diagnostics during the experiment. The good agreement between the two diagnostics, in terms of energy cut-offs and number of protons per sr, confirms the reliability of the analysis procedure. 
Such TOF detectors offer high prospects for the diagnosis of high-repetition rate (up to 10 Hz) accelerated protons and, with a suitable, automated procedure, could provide information on the shot-to-shot energy distribution and flux in real-time. In this perspective, a database containing the results of simulations for a given detector type and energy intervals, would be a key requirement for a fast, automated signal analysis.

\section{ACKNOWLEDGEMENTS}

This work has been supported by the ELIMED activities supported by the V committee of INFN (Italian Institute for Nuclear Physics), by the MIUR (Italian Ministry of Education, Research and University). 
The results of the  Project LQ1606 were obtained with the financial support of the Ministry of Education, Youth and Sports as part of targeted support from the National Programme of Sustainability  II.
Supported by the project Advanced research using high intensity laser produced photons and particles (CZ.02.1.01/0.0/0.0/16$\_$019/0000789) from European Regional Development Fund (ADONIS).

\section*{References}


\end{document}